\documentclass[letterpaper, 10 pt, conference]{ieeeconf}  

\IEEEoverridecommandlockouts                              

\usepackage{graphicx}
\usepackage{comment}
\usepackage{amsmath}
\usepackage{booktabs}

\title{\LARGE \bf
A Semi-Markov Chain Approach to Modeling Respiratory Patterns Prior to Extubation in Preterm Infants*
}

\author{Charles C. Onu$^{1\textsuperscript{+}}$, Lara J. Kanbar$^{2\textsuperscript{+}}$, Wissam Shalish$^3$, Karen A. Brown$^{4}$, Guilherme M. Sant'Anna$^3$, \\ Robert E. Kearney$^2$ and Doina Precup$^{1}$
\thanks{*Research supported in part by the Canadian Institutes of Health Research. The work of L. Kanbar was supported in part by the Natural Sciences and Engineering Research Council of Canada. K. Brown was supported in part by the Queen Elizabeth Hospital of Montreal Foundation Chair in Pediatric Anesthesia. }
\thanks{+Equal contribution from both authors. }
\thanks{$^{1}$C. C. Onu and D. Precup are with the School of Computer Science, McGill University, Montreal, QC H3A 2B4, Canada. (e-mail: { charles.onu@mail.mcgill.ca; dprecup@cs.mcgill.ca)}}%
\thanks{$^{2}$L. J. Kanbar and R. E. Kearney are with the department of Biomedical Engineering, McGill University, Montreal, QC H3A 2B4, Canada (e-mail: { lara.kanbar@mail.mcgill.ca; robert.kearney@ mcgill.ca)}}%
\thanks{$^{3}$W. Shalish and G. M. Sant'Anna are with the department of Neonatology, McGill University, Montreal, QC H3A 2B4, Canada.}%
\thanks{$^{4}$K. Brown is with the department of Anesthesia, McGill University Health Center, Montreal, QC H3A 2B4, Canada.}%
}

\begin{document}

\maketitle
\thispagestyle{empty}
\pagestyle{empty}

\begin{abstract}

After birth, extremely preterm infants often require specialized respiratory management in the form of invasive mechanical ventilation (IMV). Protracted IMV is associated with detrimental outcomes and morbidities. Premature extubation, on the other hand, would necessitate reintubation which is risky, technically challenging and could further lead to lung injury or disease. We present an approach to modeling respiratory patterns of infants who succeeded extubation and those who required reintubation which relies on Markov models. We compare the use of traditional Markov chains to semi-Markov models which emphasize cross-pattern transitions and timing information, and to multi-chain Markov models which can concisely represent non-stationarity in respiratory behavior over time. The models we developed expose specific, unique similarities as well as vital differences between the two populations.

\end{abstract}

\section{INTRODUCTION}

The respiratory management of extremely preterm infants (gestational age $\le$ 28 weeks) is specialized and challenging due to their immature lungs and inability to maintain functional breathing. Despite advances in neonatal care and noninvasive ventilation, these infants are a high-risk population that often requires invasive mechanical ventilation (IMV) during the beginning of life \cite{Walsh2005, Keszler2015}.  

Prolonged IMV is associated with several short-term and long-term morbidities, such as pneumonia, lung injury, bronchopulmonary dysplasia, and neuro-developmental problems \cite{Joseph2015,AlMandarifetalneonatal}. Therefore, clinicians emphasize the need for prompt extubation using ventilatory settings, blood gases, hemodynamic stability and clinical experience to determine readiness \cite{AlMandarifetalneonatal}.

Unfortunately, extubation failure rates in this population are high and variable, ranging from 20\% up to 70\% depending on the failure criteria and the time frame used for definition \cite{GiacconeF124}. Reintubation may be challenging and has been identified as an independent risk factor for increased morbidities and death \cite{Manley201645}. The Automated Prediction of Extubation Readiness (APEX) research project aims to quantify cardiorespiratory behavior and evaluate its association with extubation outcomes, with the goal of assisting clinicians in the extubation decision. 

As part of the APEX study, we have previously examined the ability of cardiorespiratory features \cite{precup2012} and clinical parameters \cite{gourdeau_feature_selection} to predict extubation readiness. Cardio-respiratory behavior can also be characterized by segmenting the physiological time series data into patterns of respiratory behavior that are mutually exclusive.  The resulting pattern sequence provides a high-level description of respiratory behavior. This paper describes an approach to model this sequence using different types of Markov models. The objective was to identify distinguishing characteristics of the infants who succeeded extubation compared to those who failed.

Markov chain modeling provides a robust probabilistic mechanism for characterizing sequential data. It has been applied with great success in several domains including in natural language processing \cite{rai2016_googlePA} and in gene pathway analysis \cite{li2012_markov_cancer}. In this paper, we first examined the distribution of the respiratory patterns. We then modeled the pattern-to-pattern transitions over time as discrete-time Markov chains. Because usual discrete-time Markov chains can be sensitive to the sampling rate of the signal, we also used semi-Markov models, which allow separate modeling of the time spent in each state, as well as the cross-pattern transitions. Lastly, we fit multi-chain Markov models to examine the potential non-stationarity of the respiratory patterns over time.

The rest of the paper is organized as follows: Section II describes the APEX study; Section III describes methodology and modeling details; Section IV reports the results of the experiments; and Section V discusses the conclusions of this work.

\section{APEX Study Design}

\subsection{Infant Population}

Eligible infants were recruited from Canada (Royal Victoria Hospital, Montreal Children's Hospital, Jewish General Hospital, Quebec) and USA (Detroit Medical Center, MI, and Women and Infants Hospital of Rhode Island, RI) with Birth Weight (BW) $\le$ 1250 g. Infants were recruited while undergoing IMV prior to their first extubation.

Infants were excluded if they had any major congenital anomalies such as heart disease, or were receiving any vasopressor or sedative drugs at the time of extubation. Ethics approval was obtained from the ethics board at each institution and data was collected after informed parental consent was obtained. 

\subsection{Data Acquisition}

Cardiorespiratory signals were acquired from infants in supine position in their incubators. Signals were collected for 60 min under Invasive Mechanical Ventilation (IMV), followed by a 5 min period of EndoTracheal Tube Continuous Positive Airway Pressure (ETT-CPAP) immediately prior to extubation. Respiratory signals were measured using Respiratory Inductance Plethysmography (RIP) bands placed around the infant’s ribcage (RCG), at the level of the nipple line, and around the infant’s abdomen (ABD), 0.5 cm above the umbilicus (Viasys\textsuperscript{\copyright}, Healthcare, USA). 

Signals were sampled at 1000 Hz using the PowerLab 16/30 analog-digital data acquisition system with a 16-bit analog-to-digital resolution (ADInstruments, Bella Vista, Australia, 2009\textsuperscript{\copyright}). Signals were anti-alias filtered at 500 Hz before acquisition. The signals were analyzed using  MATLAB\textsuperscript{TM} (The MathWorks Inc.).

The outcome of the extubation was also recorded. A failed extubation outcome was defined as the need for reintubation within 7 days of extubation.

\subsection{Respiratory Patterns}
\label{section_respiratory_patterns}

\begin{figure}[t]
   \centering
   \includegraphics[width=0.5\textwidth]{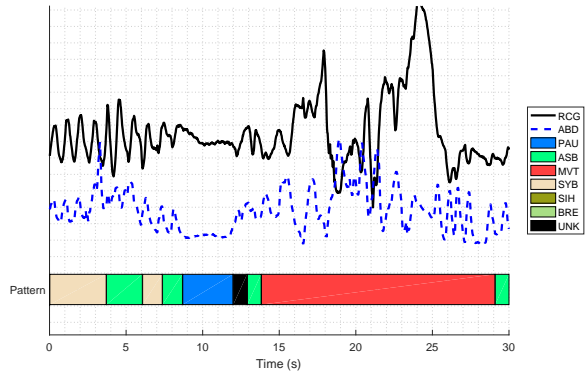}
   \caption{Example of a RCG and ABD signal segment and the corresponding respiratory patterns computed by AUREA.}
   \label{fig_RIPPatternExample}
 \end{figure}

RIP signals were analyzed using Automated Unsupervised Respiratory Event Analysis (AUREA), which extracts sample-by-sample metrics of respiratory power, synchrony between RCG and ABD, and movement artifact \cite{aurearef}. AUREA uses k-means clustering to assign each sample to one of the following respiratory patterns:
\begin{itemize}
\item Pause (PAU): A cessation of breathing.
\item Synchronous Breathing (SYB): RCG and ABD are in phase.
\item Asynchronous Breathing (ASB): RCG and ABD are out of phase.
\item Movement Artifact (MVT): Associated with infant moving or nurse handling.
\item Unknown (UNK): Ambiguous patterns not belonging to any other pattern category.
\end{itemize}
AUREA provides repeatable results with no human intervention. An example of RIP signals and corresponding patterns assigned by AUREA to the different samples is shown in Fig. \ref{fig_RIPPatternExample}. 

\section{Methodology}
The objective of this work was to model the sequence of respiratory patterns for the 5 min period of ETT-CPAP. Separate models were developed for infants who succeeded or failed extubation. 

In the following subsections, we describe our initial analysis of the pattern distributions, which motivated the choice of Markov modeling. We then describe the three different types of Markov models we applied, based on the observed data characteristics.
 
\subsection{Analysis of Pattern Distributions}
\label{section_analysis_pattern_distr}

We examined the distribution of pattern durations (for PAU, SYB, ASB, MVT, and UNK) to determine if there are marked differences over time during the ETT-CPAP period. To do this, the ETT-CPAP respiratory pattern signal is split into two equal-time segments for each patient. The length of each segment of a particular pattern was recorded. The total occurrences for each outcome group were used to plot a probability histogram. The result is a probability density histogram of the pattern durations within the first or second half of the ETT-CPAP, for either failure or success patients. 

\subsection{Fitting a Discrete-time Markov Chain}
\label{section_fit_stationary_markov}
Problems which can be phrased as transitions among discrete states over time can be modeled using Markov chains \cite{markov}. A Markov chain model consists of a  set of states $S$ (in our case, the 5 respiratory patterns); the distribution of initial states; and the transition matrix $T$, which holds the probabilities of all possible state-to-state transitions. At each times step $t=1, 2, ...$ the system is in some state, and transitions to a next states according to $T$. In a homogeneous Markov chain, the probability of transitioning from state $i$ to state $j$ is independent of the time step $t$, as well as of the history preceding state $i$:
\begin{equation} \label{eq_transition_prob}
T_{i,j} = P(X_t = j|X_{t-1} = i), \forall t, \quad where \  i,j \in S 
\end{equation}
 This is known as the Markov property \cite{markov}, which makes such models tractable by removing the need to compute full conditionals $P(X_t|X_{t-1},X_{t-2},...,X_1)$.
Empirically, the transition probabilities can be computed using maximum likelihood estimation:
\begin{equation} \label{eq_transition_prob_ml}
\hat{T}_{i,j} = \frac{n_{ij}}{\sum_j{n_{ij}}} \quad \forall \  i,j \in S 
\end{equation}
where $n_{ij}$ is number of times the transition from state $i$ to state $j$ was observed in the data.

Note that in our data, the distribution of initial states is unreliable due to infant and device handling at the beginning of data collection episodes, so in fact we did not include it in the model.

\subsection{Semi-Markov Model}
\label{section_semi_markov}

Semi-Markov models are different from standard Markov chains in that self-transitions, i.e. transitions from a state to itself, are collapsed. Instead, each state has a {\em dwell time distribution}, which models the duration spent in the state until a transition out of the state occurs, coupled with a probability distribution of transitioning to {\em other} states. The latter results in a chain where every transition results in a state that differs from its predecessor. 

Using this framework is useful for two reasons. First, it increases the resolution of the off-diagonal elements of the transition matrix $T$, especially when self-transitions are extremely likely (which is the case in our data, as will be shown in Sec.~\ref{section_experiments_results}). Secondly, if the sampling rate of the data were to change, the simple Markov chain model would drastically change, whereas the semi-Markov model would not be significantly affected. For example, if the sampling rate were to double, assuming that this does not affect the state labeling, the self-transitions of a Markov chain that works at the sampling rate would roughly double; in contrast, the semi-Markov chain would still have the same dwell time distribution and same probability of transitioning from a state to its successors.

\subsection{Multi-Chain Semi-Markov Model}
\label{section_nonstationary_markov}

Multi-chain Markov models can be used to examine non-stationarity over time in sequential data. Separate Markov chains are fit to different, non-overlapping time segments of the data. The differences in the models can be used to identify and quantify non-stationarity. We fit two semi-Markov models to the first and second half of the 5-minute respiratory sequence of patterns, partly as a soft start step, but also because modeling on a finer temporal scale (e.g. minute-by-minute) would leave too few samples for robust modeling, especially given that certain patterns (states) have low overall frequency.

\subsection{Kullback-Leibler (KL) Divergence}
The Kullback-Leibler (KL) divergence is a measure of the deviation of one probability distribution from another. The KL-divergence of a distribution P from another Q is defined as:
\begin{equation}
D_{KL}(P||Q) = \sum_i P(i) log\frac{P(i)}{Q(i)}
\end{equation}
Th KL divergence is 0 if the two distributions are identical, and it goes to $\infty$ if $Q$ does not have support for the whole domain of $P$.

Note that the KL-divergence is non-symmetric: $D_{KL}(P||Q) \ne D_{KL}(Q||P)$, which is not desirable in our application. Hence, we apply symmetrized KL-divergence to compare distributions over pattern transitions, defined as:
\begin{equation}
\begin{aligned}
\label{eq_symmetric_kl}
D_{KLS}(P||Q) = D_{KL}(P||Q) + D_{KL}(Q||P) \\ = \sum_i (P(i) - Q(i)) log\frac{P(i)}{Q(i)}
\end{aligned}
\end{equation}

\section{Experiments and Results}
\label{section_experiments_results}

At the time of this work, there were a total of 186 patients in our database: 136 extubation successes and 50 extubation failures.

\subsection{Analysis of Pattern Durations}

\begin{figure}[t]
   \centering
   \includegraphics[width=0.4\textwidth]{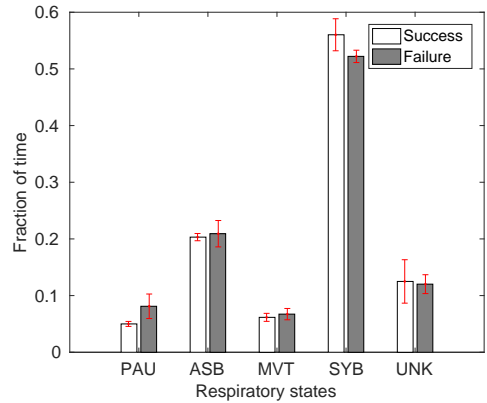}
   \caption{Fraction of time spent in a pattern during ETT-CPAP for Success and Failure patients (error bars represent standard deviations obtained by bootstrapping).}
   \label{fig_time_spent_in_state}
 \end{figure}

Fig. \ref{fig_time_spent_in_state} shows the mean fraction of ETT-CPAP time spent in each of the five respiratory patterns for the success and failure groups, with error bars to display the standard deviation. Note that failure patients spent significantly longer time in the Pause pattern, and less time in Synchronous Breathing.

\begin{figure}[t]
   \centering
   \includegraphics[width=0.5\textwidth]{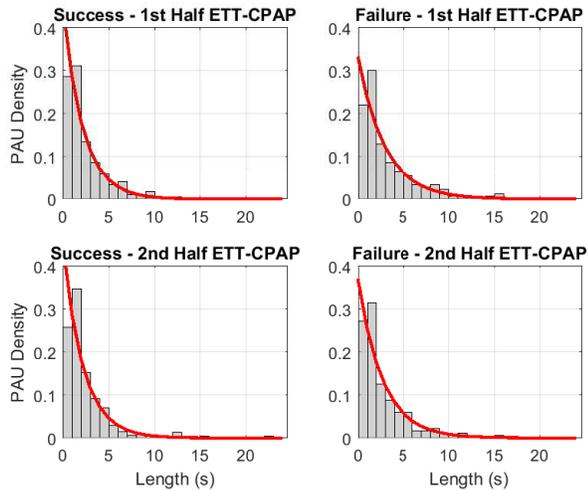}
   \caption{Pause Density Distributions across the first half and second half of ETT-CPAP. The best exponential fit (in red) for the tail of the distribution had a mean of 3s and 2.7s in the 1st and 2nd half of the ETT-CPAP for the failed extubation patients. For the success patients, it was a 2.2s mean exponential fit for both halves of ETT-CPAP.}
   \label{fig_pause_state_dist}
 \end{figure}

We also examined and fit parametric models to the histograms of durations of different respiratory patterns. Fig. \ref{fig_pause_state_dist} shows the density histograms of the Pause pattern in the success and failure groups, fit separately from the two parts of the ETT-CPAP. We also show an approximate fit using exponential distributions. The means of these distributions are different (2.2s in the success population, 3s and 2.7s in the failure population). 

\subsection{Discrete-time Markov chains}

Discrete-time Markov chains were fit to the sequence of respiratory patterns obtained during ETT-CPAP for the success and failure populations. The transition matrices (not shown) had strong and equal diagonal elements (or self-transitions). In particular, the likelihood of going from every state to itself in both the success and failure population was the same: 0.99. As a result, the numerical resolution in the off-diagonal elements (cross-pattern transitions) was very low (close to 0). Hence, the symmetric KL-divergence between the transition probabilities was close to 0 (0.001).

\subsection{Stationary Semi-Markov Model}

We fit semi-Markov models to better visualize cross-state transitions. Tables \ref{table_state_transition_success} and \ref{table_state_transition_failure} show the transition matrices for the success and failure populations, respectively. Given each start state (each row), the transition that is most likely is highlighted in bold font. It was observed that in both populations, the most probable transitions were the same for all except the Pause and Movement Artifact states. 
It should be noted, however, that even in the cases (Asynchrony, Synchrony and Unknown) where the the most probable transitions were same in both groups, the actual probabilities of these transitions differed. The symmetric KL-divergence of 0.27 also indicated a much larger difference than when using  standard Markov models. 

The dwell time distributions of the semi-Markov models were fit and are summarised in Table \ref{table_dwell_dist}. In each respiratory pattern, the {\em distribution type} which best fit the dwell times based on the Bayesian Information Criterion BIC \cite{schwarz1978estimating} was found to be same for the success and failure population. However, the values of the parameters of these distributions differed.

\subsection{Experiment 3: Multi-chain Semi-Markov model}
The 5 min ETT-CPAP data was split into 2 halves for each patient. One semi-Markov chain was fit to the first 2.5 min and another to the second. The symmetric KL divergence between the models for the success and failure groups showed an increase from 0.30 in the first half to 0.37 in the second half.



\begin{table}[b]
\caption{Respiratory state transition probabilities for the Success population modeled as a Semi-Markov chain}
\label{table_state_transition_success}
\centering 
{
\begin{tabular}{cccccc}
\toprule
 & \textbf{PAU} & \textbf{ASB} & \textbf{MVT} & \textbf{SYB} & \textbf{UNK}\\
\hline
\textbf{PAU} & 0  &  0.27   & 0.09  &  0.26   & \textbf{0.38}\\
\textbf{ASB} & 0.10  &       0  & 0.16  &   0.29  &   \textbf{0.45} \\
\textbf{MVT} &0.12  &   0.32       &  0  & \textbf{0.43}  &   0.14  \\
\textbf{SYB} & 0.06  &   0.25  &   0.15     &    0  &  \textbf{0.54} \\
\textbf{UNK} & 0.13  &0.28 &   0.04    & \textbf{0.55}      &    0 \\
\bottomrule
\end{tabular}
}
\end{table}

\begin{table}[b]
\caption{Respiratory State Transition probabilities for Failure population modeled as semi-Markov chain}
\label{table_state_transition_failure}
\centering 
{
\begin{tabular}{cccccc}
\toprule
 & \textbf{PAU} & \textbf{ASB} & \textbf{MVT} & \textbf{SYB} & \textbf{UNK}\\
\hline
\textbf{PAU} & 0  &  0.28   & 0.06  &  \textbf{0.39}   & 0.28\\
\textbf{ASB} & 0.12   &       0  &  0.21  &   0.28  &   \textbf{0.40} \\
\textbf{MVT} &0.17  &   \textbf{0.41}        &  0   & 0.32  &   0.09  \\
\textbf{SYB} & 0.14  &   0.21  &   0.14    &    0  &  \textbf{0.52} \\
\textbf{UNK} & 0.15  &0.30 &   0.03    & \textbf{0.52}      &    0\\
\bottomrule
\end{tabular}
}
\end{table}

\begin{table}[t]
\centering 
\caption{The type and parameters of the distributions of best fit to the dwell (or sojourn) times in each respiratory pattern for success and failure patients.}
\label{table_dwell_dist}
\begin{tabular}{lcc}
\toprule
 & \textbf{Success} & \textbf{Failure} \\
\hline
\textbf{Pause} & Exponential  &  Exponential   \\
 & \scriptsize{$\mu=2.51$} &  \scriptsize{$\mu = 2.94$}   \\
\cmidrule{2-3}
\textbf{Asynchrony} & GeneralizedExtremeValue & GeneralizedExtremeValue \\ 
& \scriptsize{k=0.63, $\sigma$ = 1.30, $\mu$=1.85}  & \scriptsize{k=0.65, $\sigma$ = 1.36, $\mu$=1.81}  \\
\cmidrule{2-3}
\textbf{Movement} & GeneralizedPareto &   GeneralizedPareto      \\
& \scriptsize{k=-0.22, $\sigma$ = 3.62}  & \scriptsize{k=-0.11, $\sigma$ = 3.31}  \\
\cmidrule{2-3}
\textbf{Synchrony} & InverseGaussian &   InverseGaussian \\
& \scriptsize{$\mu$ =8.61, $\lambda$ = 3.61} & \scriptsize{$\mu$ =7.83, $\lambda$ = 3.41} \\
\cmidrule{2-3}
\textbf{Unknown} & GeneralizedPareto &   GeneralizedPareto      \\
& \scriptsize{k=-0.07, $\sigma$ = 2.07}  & \scriptsize{k=-0.10, $\sigma$ = 2.05}  \\
\bottomrule
\end{tabular}
\end{table}

\section{Discussion}

The experiments presented highlight interesting differences between extremely premature infants that succeed and those who fail extubation, which are apparent during ETT-CPAP periods prior to extubation. Infants who go on to fail extubation spend a longer portion of the ETT-CPAP period in Pause and less time in Synchronous Breathing than successfully extubated patients, as seen in Figs. \ref{fig_time_spent_in_state} and \ref{fig_pause_state_dist}. 


The work also highlights the modeling power of Markov models for this type of data.  While the standard Markov chains provided models with too many self-state transitions, the use of semi-Markov models allowed modeling of cross-pattern state transitions, showing that the transition probabilities between success and failure groups were indeed different. This was seen in Tables \ref{table_state_transition_success} and \ref{table_state_transition_failure}. We also observed similarities between the populations: the most likely next respiratory state from the breathing (Synchronous and Asynchronous) and Unknown patterns were the same in both groups. 

Modeling of the dwell time distributions also showed important similarities as well as differences between the two groups. Patients who fail extubation had longer occurrences of Pauses. The dwell time in both groups followed the same distribution type in each pattern, suggesting an underlying consistency in infant respiratory behavior that is unaltered by the extubation outcome.

The multi-chain semi-Markov model presents evidence of non-stationarity in the distribution of estimated transition probabilities over time: the transition probabilities of the two groups diverged from the first to the second half of ETT-CPAP, as previously demonstrated in adults \cite{orini2008}. 

A Markov chain model encodes knowledge about state transitions and how frequently they occur. It makes the simplifying assumption that likelihoods of state transitions are independent of time, but as we have shown, non-stationarity in state transitions can be captured by fitting separate models to different time segments. The Markov models we developed based on patterns identified from RIP signals show potential for further use in  probabilistic prediction of extubation readiness. We are currently carrying out work in this direction.

The semi-Markov models can also be used to simulate breathing behavior of a patient, by sampling respiratory states from the fitted transition model, and dwell times from the timing distribution associated with each state. This allows understanding possible trends of breathing state evolution for the two populations, which can also be used to provide a measure of variability in the trajectories. 

The high probability of transitioning from SYB and ASB to UNK (Tables \ref{table_state_transition_success}, \ref{table_state_transition_failure}) raises questions about the Unknown pattern. In particular, AUREA was validated in older infants who had surgery, during the post-operative period. In extremely preterm infants, the rib cage is highly compliant and as the diaphragm shortens, there is very little expansion of the rib cage with substantial increase in the motion of the abdominal wall \cite{bryan2011}. Thus, we suspect that some of the UNK patterns found in the APEX study data may be related to predominant abdominal breathing with low-amplitude ribcage movements, which is characteristic of this population. 

In conclusion, modeling results from this work give novel insights into the respiratory behavior of extremely preterm newborns in the period prior to extubation. The ultimate goal of this research is to build classifiers that can effectively determine whether or not an infant is ready for extubation. This work highlights several relevant methods that constitute a good basis for the future work of building these classifiers.


\addtolength{\textheight}{-12cm}   





\bibliographystyle{unsrt}
\bibliography{mainbib}

\begin{thebibliography}{10}

\bibitem{Walsh2005}
Michele~C. Walsh, Brenda~H. Morris, Lisa~A. Wrage, Betty~R. Vohr, W.~Kenneth
  Poole, Jon~E. Tyson, Linda~L. Wright, Richard~A. Ehrenkranz, Barbara~J.
  Stoll, and Avroy~A. Fanaroff.
\newblock Extremely low birthweight neonates with protracted ventilation:
  Mortality and 18-month neurodevelopmental outcomes.
\newblock {\em The Journal of Pediatrics}, 146(6):798 -- 804, 2005.

\bibitem{Keszler2015}
Martin Keszler and Guilherme Sant’Anna.
\newblock Mechanical ventilation and bronchopulmonary dysplasia.
\newblock {\em Clinics in Perinatology}, 42(4):781 -- 796, 2015.
\newblock Bronchopulmonary Dysplasia: An Update.

\bibitem{Joseph2015}
Rachel~A. Joseph.
\newblock Prolonged mechanical ventilation: Challenges to nurses and outcome in
  extremely preterm babies.
\newblock {\em Critical Care Nurse}, 35(4):58--66, 2015.

\bibitem{AlMandarifetalneonatal}
H~Al-Mandari, W~Shalish, E~Dempsey, M~Keszler, P~G Davis, and
  G~Sant{\textquoteright}Anna.
\newblock International survey on periextubation practices in extremely preterm
  infants.
\newblock {\em Archives of Disease in Childhood - Fetal and Neonatal Edition},
  2015.

\bibitem{GiacconeF124}
Annie Giaccone, Erik Jensen, Peter Davis, and Barbara Schmidt.
\newblock Definitions of extubation success in very premature infants: a
  systematic review.
\newblock {\em Archives of Disease in Childhood - Fetal and Neonatal Edition},
  99(2):F124--F127, 2014.

\bibitem{Manley201645}
Brett~J. Manley, Lex~W. Doyle, Louise~S. Owen, and Peter~G. Davis.
\newblock Extubating extremely preterm infants: Predictors of success and
  outcomes following failure.
\newblock {\em The Journal of Pediatrics}, 173:45 -- 49, 2016.

\bibitem{precup2012}
D.~Precup, C.~A. Robles-Rubio, K.~A. Brown, L.~Kanbar, J.~Kaczmarek, S.~Chawla,
  G.~M. Sant'Anna, and R.~E. Kearney.
\newblock Prediction of extubation readiness in extreme preterm infants based
  on measures of cardiorespiratory variability.
\newblock In {\em 2012 Annual International Conference of the IEEE Engineering
  in Medicine and Biology Society}, pages 5630--5633, Aug 2012.

\bibitem{gourdeau_feature_selection}
P.~Gourdeau, L.~Kanbar, W.~Shalish, G.~Sant'Anna, R.~Kearney, and D.~Precup.
\newblock Feature selection and oversampling in analysis of clinical data for
  extubation readiness in extreme preterm infants.
\newblock In {\em 2015 37th Annual International Conference of the IEEE
  Engineering in Medicine and Biology Society (EMBC)}, pages 4427--4430, Aug
  2015.

\bibitem{rai2016_googlePA}
Prerna Rai and Arvind Lal.
\newblock Google pagerank algorithm: Markov chain model and hidden markov
  model.
\newblock 2016.

\bibitem{li2012_markov_cancer}
D.~Li and H.~Q. Wang.
\newblock A markov chain model-based method for cancer classification.
\newblock In {\em 2012 8th International Conference on Natural Computation},
  pages 1064--1068, May 2012.

\bibitem{aurearef}
C.~A. Robles-Rubio, K.~A. Brown, and R.~E. Kearney.
\newblock Automated unsupervised respiratory event analysis.
\newblock In {\em 2011 Annual International Conference of the IEEE Engineering
  in Medicine and Biology Society}, pages 3201--3204, Aug 2011.

\bibitem{markov}
Andrei~Andreevich Markov.
\newblock Extension of the law of large numbers to dependent quantities (in
  russian).
\newblock (2nd Ser.):135--156, 1906.

\bibitem{schwarz1978estimating}
Gideon Schwarz et~al.
\newblock Estimating the dimension of a model.
\newblock {\em The annals of statistics}, 6(2):461--464, 1978.

\bibitem{orini2008}
M.~Orini, B.~F. Giraldo, R.~Bailon, M.~Vallverdu, L.~Mainardi, S.~Benito,
  I.~Diaz, and P.~Caminal.
\newblock Time-frequency analysis of cardiac and respiratory parameters for the
  prediction of ventilator weaning.
\newblock In {\em 2008 30th Annual International Conference of the IEEE
  Engineering in Medicine and Biology Society}, pages 2793--2796, Aug 2008.

\bibitem{bryan2011}
A.~Charles Bryan, Glenn Bowes, and John~E. Maloney.
\newblock {\em Control of Breathing in the Fetus and the Newborn}.
\newblock John Wiley \& Sons, Inc., 2011.

\end{thebibliography}

\end{document}